# Memristor-based Synaptic Sampling Machines

I. Dolzhikova[1], K. Salama[2], V. Kizheppatt[1] and A. James[1]

[1]Nazarbayev University, Astana, Kazakhstan, email: ifedorova@nu.edu.kz, apj@ieee.org

[2]King Abdullah University of Science and Technology, Thuwal, Makkah Province, Saudi Arabia

*Abstract*— Synaptic Sampling Machine (SSM) is a type of neural network model that considers biological unreliability of the synapses. We propose the circuit design of the SSM neural network which is realized through the memristive-CMOS crossbar structure with the synaptic sampling cell (SSC) being used as a basic stochastic unit. The increase in the edge computing devices in the Internet of things era, drives the need for hardware acceleration for data processing and computing. The computational considerations of the processing speed and possibility for the real-time realization pushes the synaptic sampling algorithm that demonstrated promising results on software for hardware implementation.

## I. Introduction

Biological neural networks are extensively studied in the past century with an aim to mimic the human intelligence to build machine having abilities of cognition, perception and consciousness. The debates on the use and development of artificial intelligence last for ages. More than four centuries ago, Rene Descartes, one of the French mathematicians, stated that humanity would not be able to produce technology that resembles the behavioral functions of human beings [1]. The reply to this statement was given in 1950 by British mathematician who asserted that one day people would achieve such a level that the machines manufactured by them would also possess the intelligence and would be used extensively in everyday life to solve different sorts of problems [1]. The progress in mimicking neural networks and systems set forward the challenge of developing on-chip systems that are low power intelligent devices to help fully autonomous thinking machines. This is one of the major challenge of our present times, to move towards computing systems that go beyond the Von-Neumann computing paradigms, overcoming the limitations of Moore's law and device scalability through unconventional computing architectures.

Learning is the most important aspect of neural networks that helps it to adapt to the changing conditions. One of the major difficulties that always constrained the process of establishing learning models for machines is the nature of nonlinearity in the learning process [2], [3]. In addition, the latest studies observed the stochasticity in the biological mechanisms. This raised the question regarding the importance of the synaptical and neuronal unreliability in the learning algorithms of the artificial neural networks (ANNs).

The Synaptic Sampling Machine (SSM) is one of the stochastic neural networks that is based on the unreliable/random synaptic connections. Few papers could be found on the analysis of the SMM algorithms and mathematical framework [4]. In this paper we investigate the idea for the hardware realization of the synaptic sampling. The architecture is based on the two-layer memristive crossbars.

This paper is structured as follows. Section II familiarizes with the related background information. Section III demonstrates the design of the hybrid CMOS-memristive crossbar based SMM. The discussion could be found in Section IV. Finally, the conclusion is presented in Section V.

## II. Background

### A. Synaptic Sampling Machines

Random variations in the neural networks could be introduced by the activity of the stochastic neurons and/or due to unreliable synapses. If such variations are added to the ANN, the stochastic neural network that considers the probabilistic effects during neural/synaptic sampling could be obtained.

One of the first models that considers the stochastic dynamics is associated with Boltzmann machine that was presented in 1980s by Hinton and Sejnowski [5]. Boltzmann machine introduces the noise to the Hopfield network as a method to avoid convergence towards spurious local minima of the energy function. Another type of stochastic neural network that might be viewed as the variation of the Restricted Boltzmann Machine (RBM) is Synaptic Sampling Machine (SSM) that was recently introduced in the paper of Neftci et al. [4]. In contrast to the RBM, where the neurons are stochastic, SSM considers the possible failure in the synaptic connections. The probability of the given neuron $u_i$ in the SSM neural network is presented by the Eqn. (1), where $\mathrm{erf}(z_i)$ is the error function of the stochastic weighted sum ($z_i$). The terms $\varepsilon_{ij}^p(t)$, $w_{ij}$ and $b_j$ in Eqn. (2) stands for the noisy element that represents the Bernoulli process with probability $p$, weight of the synapse that connects $i$-th and $j$-th neurons, and bias, respectively.

$$P(u_i = 1) = \tfrac{1}{2}(1 + \mathrm{erf}(z_i)) \quad (1)$$

$$z_j(t+1) = \sum_{i=1}^{N} \varepsilon_{ij}^p(t) u_i(t) w_{ij} + b_j \quad (2)$$

Table 1 demonstrates the available research works on the hardware implementation of synaptic failure and the examples of their application. Stochastic synaptic connections in [6]-[9] utilizes AER. Corradi et al. [8] presented the design of system that applies the NEF's principles. The proposed prototype implemented SRAMs for synaptic weight adjustment and FPGA board for address mapping. These changes play crucial role in resulting of more precise synaptic weights and faster runtime of experiment. Furthermore, FET's are used to eliminate charge pumping effect, therefore, the noise related to that was reduced. The prototype was tested for real-time computation of mathematical functions. Authors come up to the conclusion that proposed system can be used for complex neuromorphic computations.

Paper of [6] presents a method to extend the limits of AER from the one-to-one connection to allow more complex implementation of neural circuits. Presented model uses IF transceiver connected to RAM and microcontroller, that altogether fabricated on a single chip. The system can be utilized in several scenarios as the feedforward and recurrent networks. Though, the practical results of image filtering problem have some noise due to charge pumping compared to the convolution and rectification, the time taken to run was notably faster than the time taken for simulation, that is 5 minutes and 2 hours respectively. Moreover, authors claim that the use of FPGA instead of microcontroller would increase further the experiment's speed.

Table I. Examples of hardware implementation of synaptic unreliability

| Specifications of the design | Scalability | Application |
|---|---|---|
| VLSI based design [6] (2001) | 1024 I&F neurons, each having 128 probability based synaptic connections | Image processing |
| Analog computation and digital communication [7] | 4000 neurons with 16 million recurrent synapses | Low-power embedded controllers |
| VLSI based design [8] | 58 I&F neurons, each having 32 programmable synapses and 8 additional bi-stable synapses | Brain inspired spike-based neural computation |
| 5.4 billion transistor based design, 4096 neurosynaptic cores [9] | 1 million spiking neurons and 256 million synapses | Multi-object detection and classification |

### B. SSM learning algorithm

The summary of the algorithm for learning inside the SSM network is presented below. As it could be seen there is a data phase and reconstruction phase. During the data phase the associations between the $i$-th and $j$-th unit that the network should learn based on the training data (information from the visible units) is generated. The associations during the data phase are also known as data expectations or positive expectations. During the reconstruction phase, the network produces the states of visible units based on the hypothesis about the hidden units. The associations during the reconstruction phase are known as reconstruction expectations or negative expectations. The data and reconstruction associations are used to update the weight value during the learning process.

```
SSM algorithm:
 1. for N in range(num_epoch):
 2.  ### Data phase
 3.  # i- visible_units
 4.  # j- hidden_units
 5.  z_j=dot(u_i, (weight_ij*epsilon_ij))+bias_i
 6.  Activ_func=1/2*(1+erf(z_j))
 7.  data_u_j=Activ_func>random.rand(num_examples,num_hidden + 1)
 8.  data_exp=dot(u_i.T, Activ_func)
 9.  ### Reconstruction phase
10.     rec_z_i=dot(data_u_j, w_ij.T*epsilon_ij.T)
10. Activ_func_rec_i=1/2*(1+erf(rec_z_i))
11. rec_u_j=dot(Activ_func_rec_i, w_ij*epsilon_ij)
12. Activ_func_rec_j=1/2*(1+erf(rec_u_j))
13. rec_exp=dot(Activ_func_rec_i.T,Activ_func_rec_j)
14. w_ij=w_ij+learn_rate*(data_exp-rec_exp)/num_examples
```

### III. MEMRISTOR-BASED SSMs

#### A. Memristor based synapses

Memristor is a promising electrical device that has a potential to mimic neurons and become the main element in the design of the circuits that imitate the human brain activity [10]-[11]. The recent publications about the memristive devices that were first introduced by Leon Chua in 1971 [10] reveal its important role in the design of the neuromorphic circuits. Moreover, there are positive results that demonstrate the ability of these devices to reproduce some of the biological features of the synapses [10]-[13]. One of the existing methods for construction of the neuromorphic circuits is memristor crossbar. According to [14] it is a promising approach for the development of high density neural systems.

Until recent time the Von Neumann hardware architecture was considered as one of the possible solutions for construction of neural networks [15]. However, the exploration of the capabilities of the circuit elements revealed a new possible solution. Referring to the neuroscience, synapses are the fundamental elements by which the neurons can transmit the signals to certain neural cells [16]. Synapses are responsible for the performance of the neural operations, such as computing, memorization and data processing [12]. The challenge of hardware realization is associated with non-linear behavior of the neurons that heavily depends on the time parameter. Thus, the main aspects that should be considered when trying to model an artificial neural network are that the synapse would be a crucial part of the processes connected with the large input data storage as well as decision making for further processing based on the history of previous events [12]. A large number of research works have been conducted in the area of the artificial neural systems trying to imitate the function of the biological neurons and synapses. In this paper we exploit the capabilities of the memristors to represent the synapse and introduce the stochastic effects.

#### B. Overall architecture of the proposed SSM neural network

One of the possible implementations of the synaptic sampling modeling is 1-transistor/1-memristor (1T1M) structure that emulates the artificial synapse [17]. The stochasticity of the transistor that could be achieved by varying the threshold voltage through the size scaling in addition to the deterministic or stochastic behaviour of the memristor is exactly what makes this structure applicable for the modeling of synaptic sampling. In this work however we demonstrate the stochastic behavior of the synapse by using the CMOS-switch that is controlled by the random number generator (RNG).

The overall architecture for the stochastic neural network that is based on the hybrid CMOS-memristive crossbar is illustrated on the Figure 1. The basic element of the architecture is the synaptic sampling cell. The structure of synaptic cell is shown on the Figure 2. Depending on the type of application, the input voltage to the crossbar $V_N$ is the normalized pixel value/intensity level within the range from 0 to 1. The output currents from each crossbar are converted to the voltage levels using activation function blocks. Current flow at each column of the crossbar is the weighted sum of the input that is affected by the stochasticity. The outputs from the second crossbar are processed through Winner-Take-All (WTA) circuit. The result from WTA might indicate the classification of the input data.

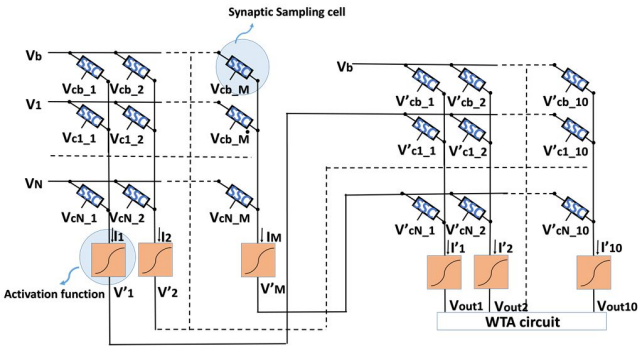

Fig. 1. The proposed architecture for SSM

The simulations were conducted using SPICE tools with the memristive device model that has threshold of $1.08V$ [14] and $50nm$ CMOS transistors ($W_{PMOS}:L_{PMOS} = 20:1$; $W_{NMOS}:L_{NMOS} = 10:1$). The activation function was implemented using Verilog A model.

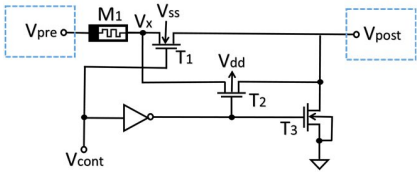

Fig. 2. Circuit design for the synaptic sampling cell

### C. Synaptic sampling cell

The synaptic cell of the proposed stochastic network consists of the memristor $M_1$ (that behaves as a synapse) and set of CMOS transistors that allows to enable/disable the synaptic connection by the control voltage $V_C$ with the probability being provided by the RNG. CMOS inverter and transistors $T_1$ and $T_2$ are used as a part of pass transistor logic to allow high and low signal to be transferred through the synapse ($M_1$) from the presynaptic signal ($V_{pre}$) to postsynaptic signal ($V_{post}$). Transistor $T_3$ is required to provide low value to $V_{post}$ when both $T_1$ and $T_2$ are OFF, in this case the synaptic connection is said to be lost. Control voltage $V_c$ is randomly generated signal with amplitudes of $1V$ or $-1V$. If $V_c$ is high the presynaptic signal is transferred to the postsynaptic terminal, otherwise the $V_{pre}$ is blocked and no signal is transferred making the $V_{post} = 0$. This idea is demonstrated

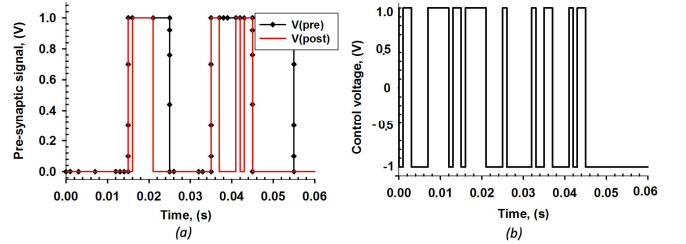

Fig. 3. Simulation results for the operating mode of SSC (a) Pre-synaptic signal and post-synaptic signal that resulted from consideration of the control voltage (b) Control voltage

through the simulation results for the single synaptic cell on Figure 3.

Random number generator having the Bernoulli distribution with the high signal having probability of $p$ could be constructed using Circular Shift Registers (CSRs), which is composed of set of D flip-flops. The output from the last flip-flop is connected to the first one, this resembles closed circle.

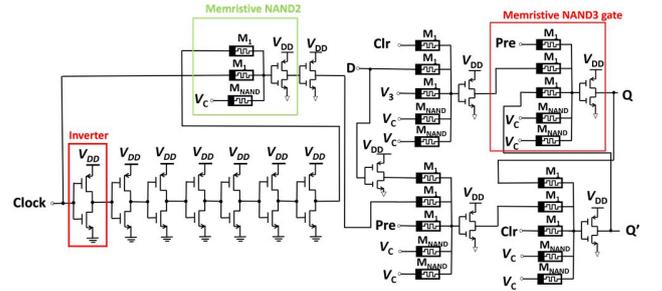

Fig. 4. CMOS-memristor based implementation of D flip-flop

Figure 4 shows the design for the memristor based D flip-flop that could be used for implementation of RNG. The comprehensive explanation of the memristive logic gates could be found in [18].

Table I summarizes the area and power calculations for the proposed design. The calculations for the control unit are presented for both CMOS and memristor based D flip-flop based implementations. In this example, the CSR consists of 10 D flip-flops. For implementation of the CMOS-based RNG, the D flip-flops in the CSR were constructed by using replacing the memristive NAND gates on the Figure 4 with the CMOS based logic elements.

Table I. Area and power calculations for the proposed design

| Circuitry | Area (um$^2$) | Power(uW) |
|---|---|---|
| SSC | 0.3525 | 0.0196 |
| Control unit (CU) | 24.45 | 3.440 |
| Memristive CU | 12.57 | 50.7 |

## IV. Discussion

### A. Proposed hardware implementation of SSM

The technological progress allowed to build the machines that are easily could be programmed such that they react to the input data and produce the output by deciding whether the statement is true or false. However, during the last few decades the progress stepped up even further. Now, the scientists and researchers are slowly moving towards the goal of teaching the machines to perform the tasks based on the obtained *knowledge* and to be more specific, to allow the machines to replicate the behavioral functionalities of the human brain.

The use of CSR for realization of the RNG allows to take the control output not only from the last D flip-flop, but also control outputs from the any other flip-flop in the chain. At a given point of time the voltages at the control outputs would be different, however they would still follow thye distribution with the probability of $p$ of having high output at every control output.

It could be noticed that benefits of memristive control unit in terms of area reduction are provided at the cost of increased power dissipation. The power consumption is increased in the memristor based implementation due to the presence of resistive path, which is absent in the CMOS design due to isolation through the ground and $V_{DD}$. In addition to that, the limitation of the memristive CSR is that it would have require longer time setup and hold time. This is due to the fact that switching time of the memristor, i.e. the time required for the memristor to change the state from low to high is around $20ns$.

## V. Conclusion

At the current state, the implementation of the proposed design involves both of the hardware and software tools, where software is used to perform learning algorithm and the hardware is responsible for the computations. In the future, the learning procedure could be also attempted to be performed on chip. Moreover, implementation of the stochastic neurons in addition to the unreliable synapses might be a good attempt for realization of the neural network that closely resembles biological functions of the brain.

### Acknowledgment

We would like to express special thanks to Rawan Nouz for the initial ideas, technical support and useful discussions on the topic of synaptic sampling. In addition to that, we thank Umesh Chand and Ulrich Buttner for guidance and support.